\documentclass[
twocolumn,
superscriptaddress,
10pt,
a4paper,
aps,
pre,
floatfix,
]{revtex4-2}

\usepackage{graphicx}
\usepackage{bm}
\usepackage{mathtools}
\usepackage{xcolor}
\usepackage{amsmath}
\usepackage{comment}
\usepackage{hyperref}
\hypersetup{backref, colorlinks=true, citecolor=blue, linkcolor=blue, urlcolor=black}

\DeclarePairedDelimiter\ceil{\lceil}{\rceil}

\newcommand{\mphi}{z}
\newcommand{\pc}{P_{z} (\rho)}
\newcommand{\ps}{\Delta_{m, \beta} (\rho)}

\begin{document}

\title{Competition between simple and complex contagion on temporal networks}

\author{Elsa Andres}
 \email{elsaandres1@gmail.com}
\affiliation{
 Department of Network and Data Science,  Central European University, Vienna, 1100, Austria
}

\author{Romualdo Pastor-Satorras}
\affiliation{
Departament de Física, Universitat Polit\`ecnica de Catalunya, Campus Nord B4, Barcelona, 08034, Spain
}

\author{Michele Starnini}
\affiliation{
Department of Engineering,
Universitat Pompeu Fabra, 
Barcelona, 08018, Spain
}
\affiliation{
CENTAI,
Torino, 10138, Italy
}

\author{Márton Karsai}
\affiliation{
 Department of Network and Data Science,  Central European University, Vienna, 1100, Austria
}
\affiliation{
 National Laboratory of Health Security, HUN-REN Alfréd Rényi Institute of Mathematics, Budapest, 1053, Hungary
}

\date{\today}

\begin{abstract}
Behavioral adoptions are influenced by peers in different ways. While some individuals may change after a single incoming influence, others need multiple cumulated attempts. These two mechanism, known as the simple and the complex contagions, often occur together in social phenomena alongside personal factors determining individual adoptions. Here we aim to identify which of these contagion mechanism dominate a spreading process propagated by time-varying interactions. 
We consider three types of spreading scenarios: ones pre-dominated by simple or complex contagion, and mixed dynamics where the dominant mechanism changes during the unfolding of the spreading process.
We propose different methods to analytically identify the transitions between these three scenarios and compare them with numerical simulations. This work offers new insights into social contagion dynamics on temporal networks, without assuming prior knowledge about individual's contagion mechanism driving their adoption decisions.
\end{abstract}

\maketitle

\emph{Introduction.}
Our personal decisions ---buying products, following trends, or adopting hobbies--- are shaped not only by our own preferences but also by the influence of others~\cite{christakis2013social,castellano2009statistical, jose2022covid, christakis2008collective},
that bias our choices to make us align with the larger group~\cite{hodas2014simple}. 
This phenomenon, known as social contagion~\cite{christakis2013social}, leads to widespread patterns of adoption across populations. 
While it has long been understood that interpersonal influence affects mass adoption~\cite{bass1969new,rogers2014diffusion}, the mechanisms behind different social contagion effects remain unclear.
At the individual level, according to the social brain hypothesis \cite{dunbar1998social}, the strength of peer influence may vary among our acquaintances depending on the intimacy level~\cite{easley2010strong}. Meanwhile, reinforcement mechanisms could also play a role, leading to adoption through repeated social stimuli~\cite{granovetter1978threshold, schelling1971dynamic,karsai2014time}.

Multiple social contagion mechanisms have been identified to explain the diversity of observed contagion patterns~\cite{christakis2013social}. 
These are often modeled as binary-state processes, where individuals (or nodes) move from a susceptible to an infected state through interactions~\cite{castellano2009statistical, gleeson2013binary}.
{Simple contagion} models, like the Susceptible-Infected (SI) dynamics~\cite{daley1965stochastic,maki1973mathematical}, assume that a single contact could be enough to trigger transmission, leading to smooth initial growth of spreading~\cite{barthelemy2004,pastor2015epidemic}.
In contrast, complex contagion models~\cite{centola2007complex} require multiple exposures before adoption occurs, once a cognitive threshold is met~\cite{granovetter1978threshold, schelling1971dynamic}.
This type of mechanism accounts for rapid cascading spreading dynamics, commonly seen with highly popular products or memes~\cite{watts2002simple}.

Contagion models often rely on structured social networks to drive their dynamics~\cite{barrat2008dynamical,watts2002simple}. 
Several topological properties, such as the egocentric network size (degree)~\cite{barrat2008dynamical, chen2022information, moreno2004dynamics}, triadic closure (clustering)~\cite{centola2010spread}, peer-wise strength of influence (link weights)~\cite{granovetter1973strength, easley2010strong,yang2012epidemic,unicomb2018threshold}, or multilayer structure~\cite{de2016physics,unicomb2019reentrant} play important roles in the spreading process.
Furthermore, the time-varying nature of social interactions is less studies in this context, while it heavily influences the contagion dynamics, as it determines the time-respecting paths along which information or influence can be disseminated.
The lifetime of links~\cite{holme2014birth}, the frequency of interactions~\cite{karsai2011small,karsai2016local}, the limited waiting times of processes at nodes~\cite{badie2022directed}, the causally related adjacent events~\cite{kivela2012multiscale}, the memory length of influence~\cite{unicomb2021dynamics,takaguchi2013bursty,karimi2013threshold}, or the heterogeneous bursty interaction dynamics~\cite{unicomb2021dynamics,takaguchi2013bursty,karsai2011small,karsai2018bursty} all have been identified important to influence spreading processes on temporal networks.

These studies typically focus on either simple or complex contagion mechanisms to explain the actual spreading phenomena.
Some aim to differentiate between these mechanisms~\cite{cencetti2023distinguishing,contreras2024infection} 
either from local knowledge only~\cite{andres2024distinguishing} or by studying their co-evolution \cite{hebert2020macroscopic}. 
In others, contagion mechanisms vary across network layers or community structures~\cite{czaplicka2016competition,diaz2022echo}, while some assign individual thresholds to nodes, specifying the number of exposures needed for adoption~\cite{karampourniotis2015impact, dodds2004universal, karsai2016local, wang2016dynamics, min2018competing}. 
The infectiousness of the spreading process and the underlying network structure often determine which mechanism dominates~\cite{horsevad2022transition}. In a case of a mixture of simple or complex adoption cases, the easier contracted simple contagion cases govern the spreading initially and they trigger in turn the complex contagion adoptions that are conditional to a fraction of peers to be infected a priori~\cite{min2018competing}. 

However, in real-world scenarios, individuals are not limited to a single contagion mechanism: they might adopt behaviors through both simple and complex contagions depending on their social context. For example, adoption might occur from a single intimate contact (simple contagion) or through repeated interactions with distant acquaintances (complex contagion). Therefore, considering the interplay between these mechanisms is pivotal to approximating more realistic social phenomena. In this paper, we propose a new approach to identify whether simple, complex, or mixed contagion processes dominate at different stages of spreading dynamics in temporal networks. We classify processes into three categories and introduce methods to analyze transitions between them based on various parameters. In our simulations an individual is not endowed with a pre-assigned contagion mechanism but, depending on the peers involved, they could adopt via both simple or complex contagions.

\emph{Model definition.} We simulate the dynamics of time-varying interactions by building on the activity-driven network (ADN) model, that provides a flexible modelling framework of temporal networks~\cite{perra2012activity,Starnini.2013}.
We consider $N$ initially disconnected nodes, each of them assigned with an activity $a_i \in [0,1]$, extracted from an arbitrary distribution $F(a)$. 
The network evolves through asynchronous iterations of $N$ microscopic time steps, in which each node is updated once on average. 
In a single microscopic time-step, of duration $\Delta t = 1/N$,  a node $i$ is selected randomly and it becomes activated with probability $a_i$~\footnote{Here we define the activity as an activation probability. Other works~\cite{Starnini.2013} consider $a_i$ as an activation rate, or probability per unit time}.
In case of activation, the node connects to $m$ randomly selected other nodes from the network,  avoiding self-loops and multi-links. 
In the following iteration, we delete all links and start the process over again.

To model a binary state contagion process~\cite{gleeson2013binary} on the top of an ADN model, we assume that each node at any time step can be in one of two mutually exclusive states: susceptible (S), thus not yet reached  by the contagion process, or infected (I) if it has been contaminated.
We set all nodes initially as susceptible and start the spreading from a small set of randomly selected infected seed nodes. Once a susceptible node is contaminated during the spreading, it remains in this state, thus the system is evolving towards a fully infected absorbing state. 
Susceptible nodes can become infected in two ways: either get contaminated through the \textit{simple contagion} mechanism, with an independent probability $\beta$ for each contact with an infected peer;
or through the \textit{complex contagion} mechanism, if the fraction of infected neighbors exceeds a pre-assigned threshold $\phi \in [0,1]$. 
The simple contagion is thus driven by a stochastic process,
while the complex contagion is fully deterministic, ruled by the threshold and the proportion of infected neighbors of each node at each time step.

\begin{figure}[tbp]
\includegraphics[width=0.95\columnwidth]{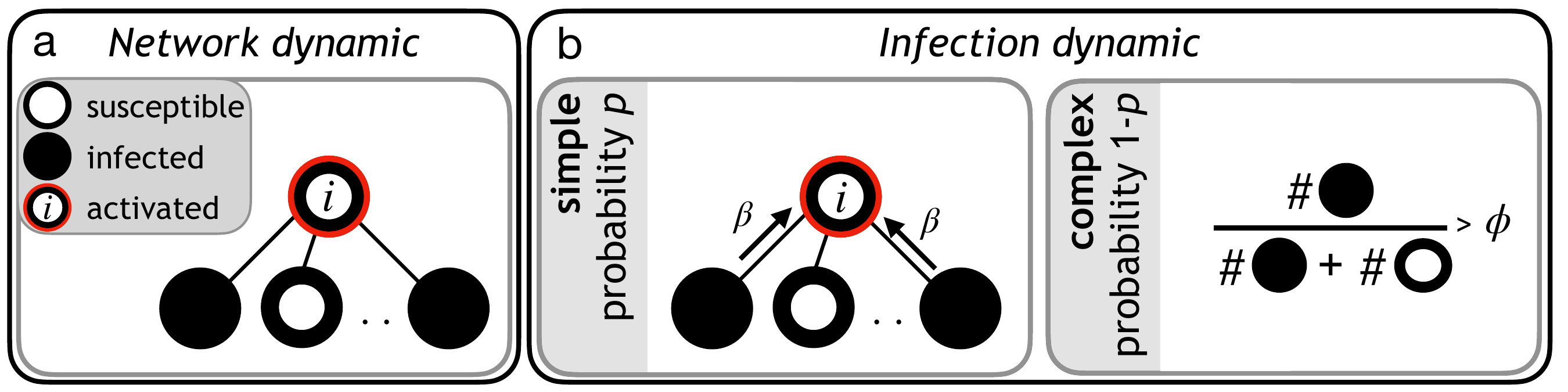}
\caption{Network and contagion dynamics. (a) A node $i$ is activated with probability $a_i$ and connects to $m$ randomly selected nodes. 
If the activated node $i$ is susceptible, with probability $p$ it follows the simple contagion mechanism (b) and gets infected by each infected neighbor with probability $\beta$. With probability $(1-p)$ it follows the complex contagion rule (c) and gets infected if the fraction of its infected neighbors is above $\phi$.}
\label{schema}
\end{figure}

We schematically summarize the network dynamics and the two contagion mechanisms in Fig. \ref{schema}. We initially infect a randomly selected $1\%$ of seed nodes. 
Subsequently, at each time-step, 
if an activated node is susceptible, it will follow the simple contagion mechanism with probability $p$, and the complex contagion one with probability $1-p$. 
In the former case, the node may get infected with probability $\beta$ from any of its infected neighbors. 
Otherwise, it gets infected if the fraction of the infected neighbors exceeds the threshold $\phi$. In practical terms, if $n_i$ is the number of infected nearest neighbors of the susceptible node $i$, complex contagion takes place if $n_i \geq z$, where $z = \ceil*{m \phi}$.
These steps are repeated until every node is in the infected state.

In the following, we denote by $I(t)$ the number of infected nodes at time $t$ and by $\rho(t) = I(t)/N$ ($\rho$ for brevity) the fraction of infected nodes (prevalence), while $\rho_0$ being this quantity at the start of the propagation.
Likewise, $\rho_s$ and $\rho_c$ indicate the fractions of nodes infected by the simple  or complex contagion, respectively.
To simplify the mathematical description, we formulate some assumptions about the propagation process.
We model a homogeneous network dynamics by assuming that all nodes have the same activity $a$. We consider the heterogeneous case in the Supplementary Material. 
Further, we study the cases of $\mphi = 2$ and $\mphi = 3$. 
The analytical treatment for more general cases with any value of $\mphi$ is reported in the Supplementary Material.

\emph{Analytical study.} To shed analytical light on the model, we consider its description in terms of a mean-field rate equation, inspired by Ref.~\cite{perra2012activity}, for the fraction of infected nodes $\rho(t)$. In our description we take into account the effects of activity and the two alternative methods of infection, either simple of complex contagion, see Supplementary Material, Section S-I.

We first consider the case where only complex contagion is present, corresponding to $p=0$. The exact rate equation cannot be easily solved analytically. Therefore, in order to get an insight of the dynamics, we focus on the short time evolution, when $\rho \ll 1$, corresponding to a leading order expansion (see Supplementary Material, Section S-II). In this regime, the density of infected nodes follows the equation
\begin{equation}
    \rho^{\mphi-1}(t) = \frac{1}{\rho_0^{1 - \mphi} - (\mphi -1) C t},
    \label{decrease_linear}
\end{equation}
where $C$ is a constant depending on $a$, $m$ and $z = \ceil*{m \phi}$. This function shows a linear decrease in time of the quantity $1/\rho^{z-1}$, and a divergence at time $t = \rho_0^{1-z}/[C(1-z)]$.  This divergence is non-physical since it appears as a higher-order approximation; adding other terms in the full equation curbs the divergence and leads to a prevalence $\rho(t) \leq 1$. However, we can interpret this apparent divergence in opposition to the behavior of pure simple contagion, in which prevalence grows exponentially in the linear approximation~\cite{barthelemy2004}. Complex contagion operates instead in cascades, in which a large fraction of nodes become infected in a very short period of time~\cite{watts2002simple}. Thus, we can identify the divergence time as the onset of the cascading behavior of complex contagion.

When both  simple and complex contagions are present ($0 < p < 1$), and focusing on the case $z=2$, the evolution of the prevalence reads
(see Supplementary Material Section  S-III)
\begin{equation}
    \frac{d \rho}{dt} = a p m \beta (\rho + B \rho^2)
    \label{rho_simpler}
\end{equation}
up to order $\rho^2$, where we have defined the constant $B = \frac{(1 - p) (m - 1)}{2 p \beta} - \frac{(m - 1) \beta}{2} - 1$.
The solution of Eq.~\eqref{rho_simpler} in terms of the initial density of infected nodes is
\begin{equation}
\label{time_increase}
    \rho(t) = \frac{1}{A e^{- a p m \beta t} - B},
\end{equation}
where $A = B + \frac{1}{\rho_0}$. The sign of $B$ determines the behavior of the prevalence: for $B<0$, $\rho(t)$ has a monotonous behavior, while for $B>0$ $\rho(t)$ shows again a divergence. Within the higher-order approximation, this can be understood at $B$ negative corresponding to simple contagion, and $B$ positive to complex contagion. The transition between those two regimes arises when $B=0$, yielding the critical value of $p$ separating the regimes:
\begin{equation}
    p_c(\beta) = \frac{m-1}{2 \beta + (1+\beta^2)(m-1)}.
    \label{eq:critica_p}
\end{equation}
If $B$ is positive, the prevalence diverges at
\begin{equation}
    t_\mathrm{casc} = \frac{1}{a p m \beta} \ln\left(\frac{A}{B} \right) ,
    \label{t_increase}
\end{equation}
which serves as a proxy of the time of onset of the cascading behavior in complex contagion.

Finally, for $B > 0$, by analyzing the evolution of Eq. \eqref{rho_simpler}, we can calculate when the first term in the rhs, corresponding to pure simple contagion, is equal to the second term, with contributions of both the simple and complex contagions. These terms are equal at the prevalence 
\begin{equation}
    \rho_\mathrm{eq} = 1/B,
    \label{eq:rho_eq_main}
\end{equation}
which corresponds to  the time:
\begin{equation}
    t_\mathrm{eq} = \frac{1}{a p m \beta} \ln\left(\frac{A}{2 B} \right) .
    \label{t_eq}
\end{equation}
At this time, complex contagion takes over the initial simple contagion dynamics, in the complex dominated regime $p < p_c(\beta)$, a phenomenon which has already been identified previously \cite{min2018competing}.

\emph{Numerical simulations.} We contrast our analytical results with numerical simulations, with parameters $N=1000$, $a=1$, $m=5$ and $\phi=0.25$ for $z=2$, and $\phi = 0.5$ for $z=3$.
We average simulations over $100$ runs. 
We fix $z=2$ (if not noted otherwise) while varying $\beta$ and $p$.

Fig.~\ref{proportion_infection_by_type} shows the contagion curves for different $\beta$ and $p$ (see Fig.~S1 in  Supplemental Material for the full parameter space). 
When $p$ is large, the dynamics is determined by simple contagion, as we can see by the fast growth of the prevalence, which is faster for larger values of $\beta$, see Fig.~\ref{proportion_infection_by_type}a and~\ref{proportion_infection_by_type}b.  
When $p$ is small, on the other hand, the complex contagion prevails, see Fig.~\ref{proportion_infection_by_type}c-d. In this regime, the prevalence grows initially slowly, followed by a sudden increase indicative of cascading. 
Interestingly, as shown in Fig.~\ref{proportion_infection_by_type}d, the $\rho_c$ cascade emerges earlier when rare ($p=0.05$) but likely transmitting ($\beta=0.99$) simple contagion events are present. 
Since the initial seeds may not be eligible to induce a complex contagion outbreak, simple contagion cases build up the necessary initial conditions to trigger complex contagion.

Fig.~\ref{proportion_infection_by_type} and Fig.~S2 (Supplemental Material)  
show that the anlytical divergence time $t_{casc}$ generally predicts quite closely the time when the infection curve $\rho(t)$ starts taking off, signaling the onset of cascades.
Nevertheless, $t_{casc}$ slightly underestimates the real outbreak time if both $\beta$ and $p$ have high values, while $t_{casc}$ is somewhat late if $\beta$ is low and $p$ is high.
Further, we note that a non-constant activity distribution can affect the propagation speed. Supplementary Material Section~S-V and Fig.~S1 show that, when activities are drawn from a power-law distribution, the epidemic spreads faster initially, but it takes longer to reach every node, which is in contrast with the case when all nodes have the same activity level.

\begin{figure}[tbp]
\centering
\includegraphics[width=0.95\columnwidth]{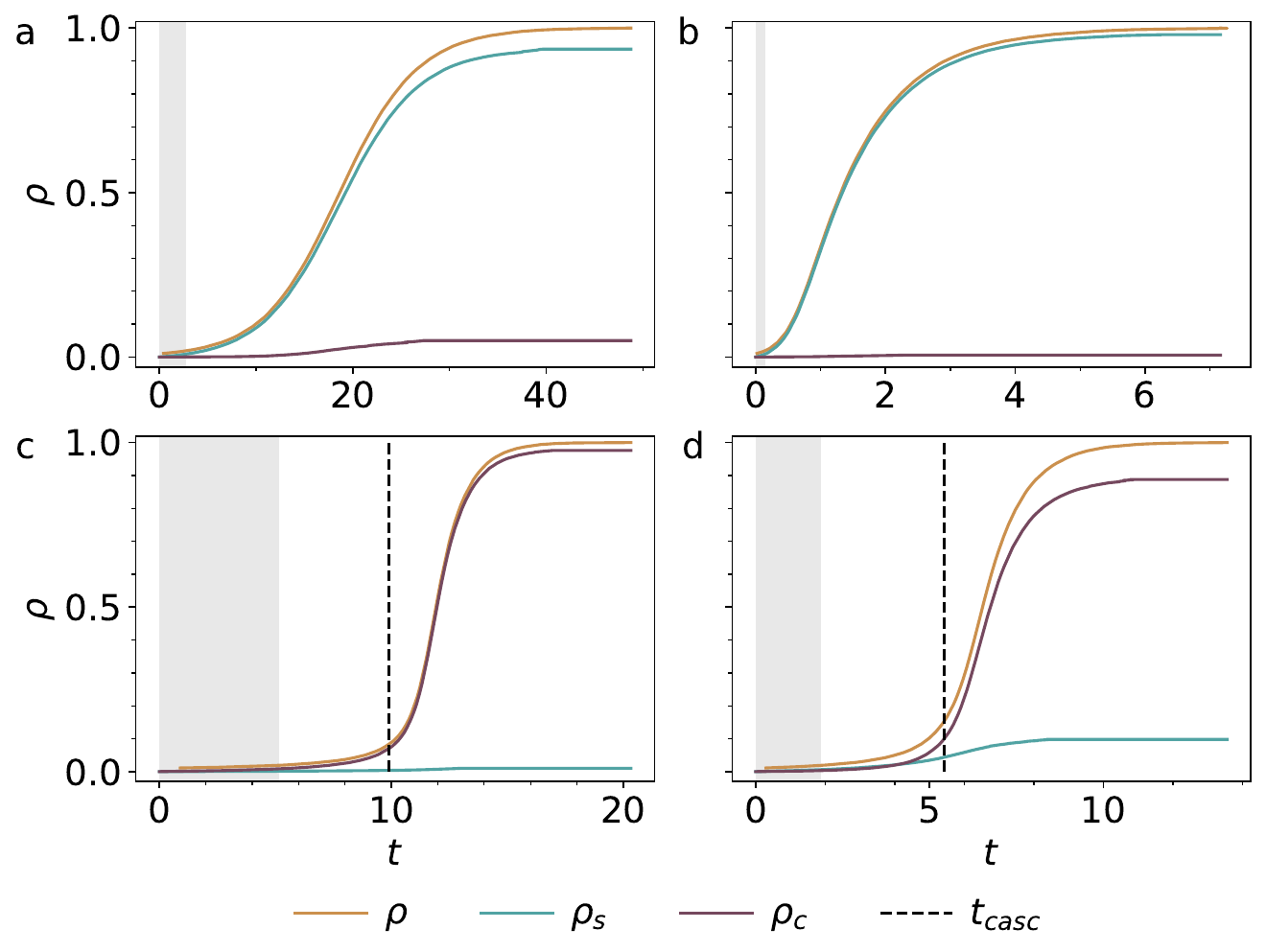}
\caption{Fraction of infected nodes $\rho$ (yellow line), proportion of nodes infected by simple $\rho_s$ (blue) and complex $\rho_c$ (purple) contagion as a function of time, for $z=2$. 
Panels show different $(\beta, p)$ parameters: a) $(0.05, 0.99)$, b) $(0.99, 0.99)$, c) $(0.05, 0.05)$, and d) $(0.99, 0.05)$. 
The simple (complex) contagion dominates the propagation when $p$ is high (low), with a minor influence from $\beta$, which can be observed on Fig. S2. 
The gray area indicates the early period of the contagion up to $t_{init}$, when $\rho=0.02$. 
Dashed vertical lines indicate $t_{casc}$, corresponding to the expected outbreak time if $B$ is positive (see Eq. \eqref{t_increase}), in panels c) and d). 
}
\label{proportion_infection_by_type}
\end{figure}

Following this qualitative analysis, we focus now on disentangling the effective mechanisms that rule the evolution of the spreading process.
As we have mentioned above, the dominant mechanism can change during the process in the mixed simple/complex scenario. We thus consider the early and late stages of the propagation separately.
The early stage corresponds up to the time $t_{init}$, at which the prevalence $\rho$ fulfills the condition $\rho=0.02$, 
while the  late stage encompasses from $t_{init}$ to the time when all nodes are infected,  $t_{end}$.
To identify the early and late contagion mechanisms we consider the ratios $\lambda=\rho_c(t_{init})/\rho_s(t_{init})$ and $\Lambda=\rho_c(t_{end})/\rho_s(t_{end})$.
If the simple (complex) contagion dominates the entire process, both quantities remain below (above) 1.
If the simple contagion dominates the early stage but the complex one takes over in the late stage, then we expect $\Lambda > 1$ and $\lambda < 1$ respectively.

Since the case $\Lambda < 1$ and $\lambda > 1$ cannot be observed (once the complex contagion is triggered, it propagates much faster than the simple one), we classify the spreading dynamics into three categories: pure simple contagion ($\Lambda < 1$ and $\lambda < 1$), pure complex contagion ($\Lambda > 1$ and $\lambda > 1$), and mixed contagion ($\Lambda > 1$ and $\lambda < 1$).
We thus expect two transitions in the parameter space $(\beta, p)$. The first, from pure simple to mixed contagion, and the second from mixed to pure complex. The first transition takes place at a threshold $p_c(\beta)$, given by Eq.~\eqref{eq:critica_p} for $z=2$, separating the phase in which simple contagion is dominating asymptotically from the phase in which complex contagion dominates at large times. In the mixed phase, while complex contagion is dominant at large times, at short times simple contagion is prevalent. The second transition separates this mixed phase from the pure complex contagion phase, in which even at short times complex contagion is predominant.

In the following, we propose two different methods to identify the transition point from mixed to pure complex phases.

\paragraph*{Method 1:} In the complex dominated phase, we measure the time $t_\mathrm{eq}$ and prevalence $\rho_\mathrm{eq} = \rho(t_\mathrm{eq})$ when the two terms in the rhs of Eq. \eqref{rho_simpler} become equal, indicating when the simple contagion term takes over the complex contagion one. These quantities can be computed analytically from Eqs.~\eqref{eq:rho_eq_main} and~\eqref{t_eq}, and numerically by evaluating in simulations when the first and second terms in the rhs of Eq. \eqref{rho_simpler} become equal. The time  $t_\mathrm{eq}$ signals the transition between a dynamics initially dominated by simple contagion and the dynamics asymptotically dominated by complex contagion. In the parameter space $(\beta, p)$, the transition from the mixed phase to the pure complex contagion phase should thus correspond to $t_\mathrm{eq}(\beta, p) = 0$ (noted $t_{eq}$ \textit{null} in the following figures), that is, when right at the initial time step the contagion is dominated by the complex mechanism.

\begin{figure}[tbp]
    \centering
    \includegraphics[width=0.95\columnwidth]{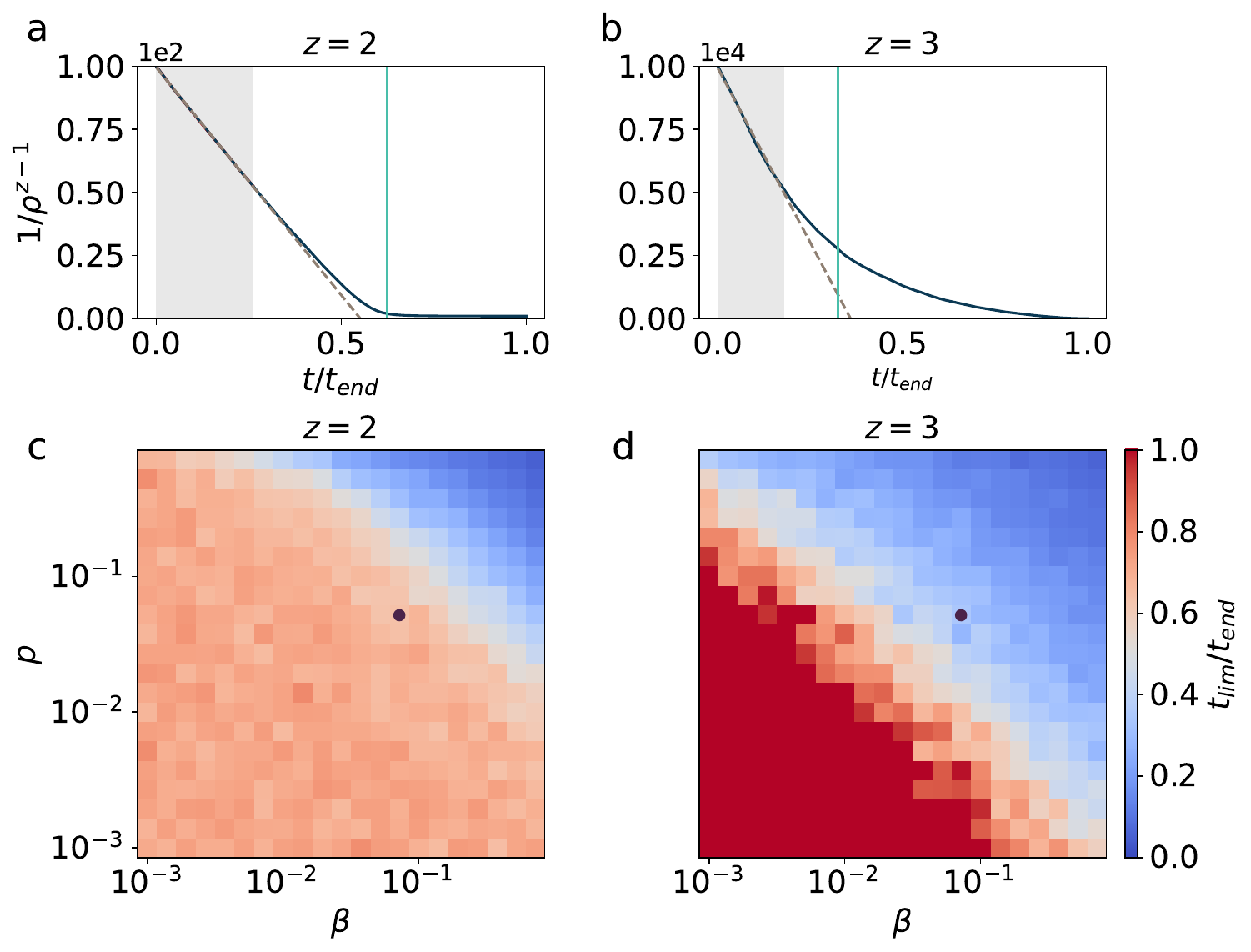}
    \caption{
    The $1/\rho^{\mphi-1}$ curve (dark blue line) obtained from simulation averages with 
    $(\beta,p)=(0.037, 0.037)$, fitted by a linear function (black dotted line) on the period before $t_{init}$ (gray area) for $z=2$ (panel a) and $z=3$ (panel b). 
    The $t_{lim}$ time point when the two curves diverge is corresponding to an $\epsilon>0.15 / \rho_0^{z-1}$ difference, indicated by a vertical light blue line. The different $t_{lim}$ obtained across the whole parameter-space $(\beta, p)$ are displayed for $z=2$ (panel c) and $z=3$ (panel d).
    $1/\rho^{\mphi-1}$ stops being linear at early times when the propagation is fast i.e., when both $\beta$ and $p$ are high, while this linearity persists longer when the contagion process is slow (low $\beta$, and high $p$ for $z=2$). 
    Simulations results were calculated as averages over 100 realisations.}
    \label{fig_linear_decreasing}
\end{figure}

\paragraph*{Method 2:} The second approach relies on the results in Eq.~\eqref{decrease_linear}, suggesting that the function $1/\rho^{\mphi-1}$ should decrease linearly with time $t$, if the spreading is governed by pure complex contagion. 
We demonstrate this behavior in Fig.~\ref{fig_linear_decreasing}a and b (for $z=2$ and $z=3$ respectively) by showing the curve $1/\rho^{\mphi-1}$ for simulations corresponding to the $(\beta,p)$ pairs indicated by the black points in Fig.~\ref{fig_linear_decreasing}c and d respectively.
In the case where both simple and complex contagions are present, looking at the short time behavior of $1/\rho^{\mphi-1}$ is also linear with $t$, see Eq.~\eqref{time_increase}, as it can be seen expanding the exponential in the denominator regardless the sign of $B$. 
However, this linear trend breaks down earlier,  since the expansion of the exponential fails sooner. As a result, the complex contagion dominates longer, as marked by a longer linear decrease. To quantify this effect, we first consider an initial regime up to small initial time $t_{init}$, in which we assume a linear behavior for $1/\rho^{\mphi-1}$. 
To take equally different values of $z$, we define  $t_{init}$ as the time when $\rho$ satisfied the condition $\frac{1}{\rho^{z-1}}/\frac{1}{\rho_0^{z-1}} = 0.5$ (if $z=2$ and $\rho_0=0.01$, this condition corresponds to $\rho=0.02$).
We then fit a linear function to $1/\rho(t)^{\mphi-1}$ in the interval $[0, t_{init}]$. Finally, we find the value $t_{lim}$ at which the linear behavior breaks down. This is defined by the time at which the function $1/\rho(t)^{\mphi-1}$ differs from the initial linear fit by a value larger or equal than $\epsilon = 0.05 / \rho_0^{z-1}$. The prevalence at this time is denoted as $\rho_\mathrm{lim}=\rho(t_\mathrm{lim})$.

In Fig.~\ref{fig_linear_decreasing} we show an example for the application of Method 2. 
Indeed, as Fig.~\ref{fig_linear_decreasing}a depicts, $t_{lim}$ takes a larger value as the $1/\rho^{\mphi-1}$ curve scales linearly over an extended regime, suggesting dominating complex contagion. On the contrary, in Fig.~\ref{fig_linear_decreasing}b, the simulations are dominated by simple contagion, thus $t_{lim}$ is small as a shorter linear scaling is observable.
We display the rescaled time $t_{lim}/t_{end}$ for an extended parameter space in Figs.~\ref{fig_linear_decreasing}c  and Figs.~\ref{fig_linear_decreasing}d for $z$ equal to $2$ and $3$, respectively.
According to these results our method can well separate the regime where the complex contagion dominates (characterized by large $t_{lim}/t_{end}$ values and  corresponding to lower values of $\beta$ and $p$), form the simple contagion dominated regime (with small values of $t_{lim}/t_{end}$ and for higher values of $\beta$ and $p$).

\begin{figure}[tbp]
    \centering
    \includegraphics[width=0.95\columnwidth]{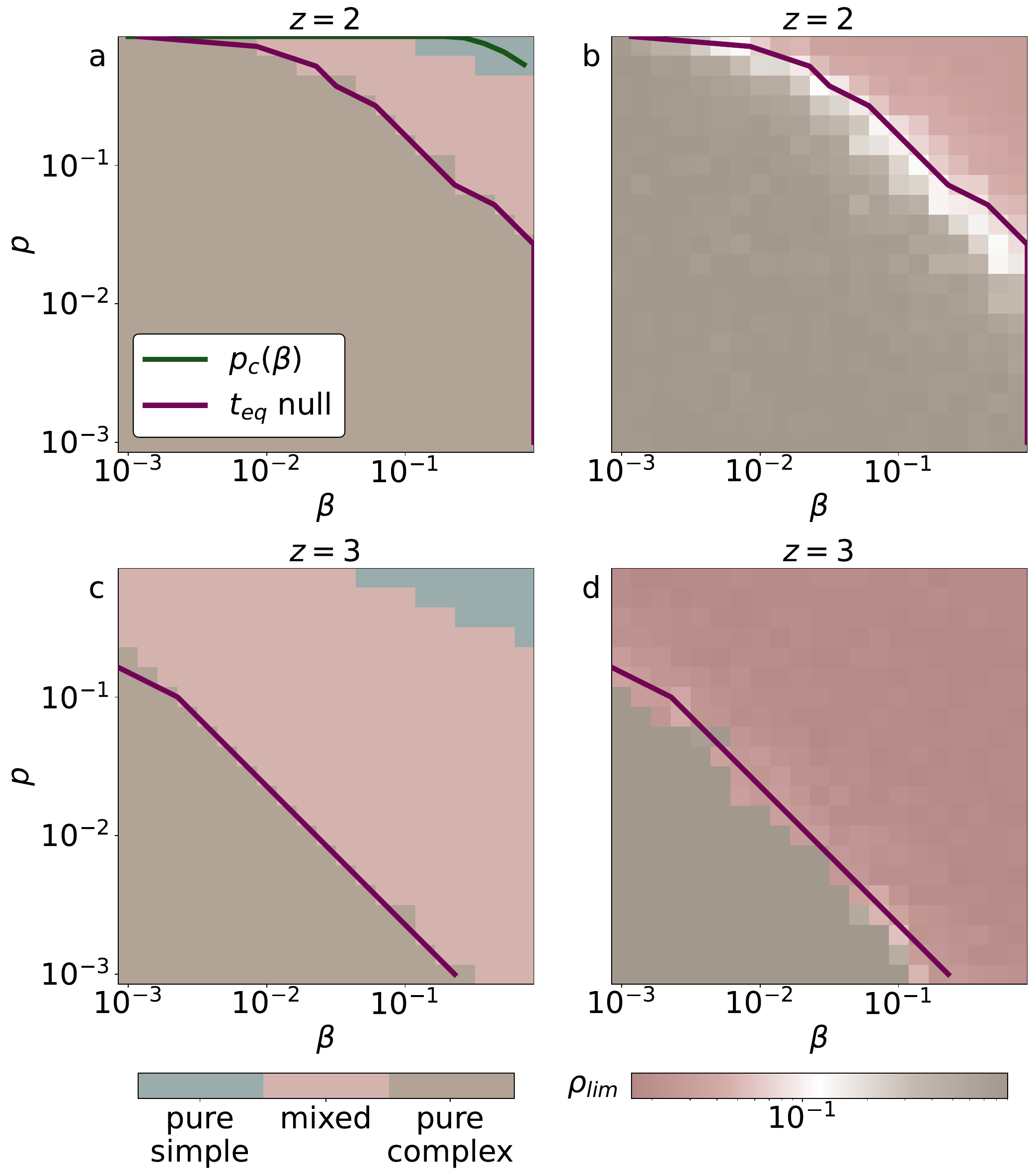}
    \caption{Areas corresponding to the three categories in the parameter-space $(\beta, p)$ on panel a for $z=2$ and on panel b for $z=3$. The quantities $p_c(\beta)$ and the limit of $t_{eq}$ null, indicated by a green and a purple line respectively, matches the transitions between those areas. The values of $\rho_{lim}$, displayed on the second column, are also marked by the transition between pure complex and mixed.}
    \label{fig_matrices_mphi_2}
\end{figure}

Fig.~\ref{fig_matrices_mphi_2} shows how the analytical and numerical methods capture the transitions from pure simple contagion to mixed, and from mixed to pure complex, in the $(\beta, p)$ parameter space. 
Figs.~\ref{fig_matrices_mphi_2}a and~\ref{fig_matrices_mphi_2}c show a phase diagram depicting the different regions in the $(\beta, p)$ space for $z=2$ and $z=3$, respectively. 
The regions are determined by comparison of the values $\lambda$ and $\Lambda$ defined above: pure simple contagion for $\Lambda, \lambda < 1$,  pure complex contagion for $\Lambda, \lambda > 1$,  and mixed contagion for $\Lambda > 1$, $\lambda < 1$. 
For $z=2$, the phase boundary between pure complex (brown area) and mixed (pink area) is well captured by Method 1, corresponding to a null $t_\mathrm{eq}$ line in red. 
The boundary between mixed and pure simple (green area) is slightly shifted from the prediction $p_c(\beta)$ (green line). In case of $z=3$, again Method 1 provides an excellent approximation for the boundary between pure complex/mixed phases, which seems to take place at smaller values of $(\beta, p)$ than for $z=2$. In this case, we do not have an analytical prediction for the boundary mixed/pure simple, leaving us alone with the results of numberical simulations. 
However, it seems to appear again for smaller  $(\beta, p)$ values. 

In Figs.~\ref{fig_matrices_mphi_2}b and~\ref{fig_matrices_mphi_2}d, we present the predictions by Method 2 for the location of the boundaries between pure complex/mixed phases for the cases of $z=2$ and $z=3$, respectively. In these plots we depict as color maps the density $\rho_\mathrm{lim}$ for the whole range of the $(\beta, p)$ parameter space. As we can see, the collapse of the boundary extracted from Method 2 with the classification made in terms of $\Lambda$ and $\lambda$ ratios is excellent, while also being in very good agreements with the prediction of Method 1 represented by $t_\mathrm{eq}$ null.

\emph{Conclusions.} In this paper, we introduced a mixed model of social contagion on temporal networks, in which nodes can be infected by either simple or complex contagion dynamics simultaneously. 
We focused on the simplest modeling scenario of a homogeneous activity-driven network with all nodes having the same activity. We tackled the solution of the model analytically using a mean-field rate equation for the total density of infected nodes as a function of time, and numerically through extensive simulations. This way we proposed two ways to differentiate between processes of different dominant contagion processes.

We identified three phases of contagion according to the two main parameters of the model, the probability of adopting by simple contagion $p$ and the infection probability of simple contagion $\beta$.
For small (large) values of $(\beta, p)$, the dynamics is ruled purely by simple (complex) contagion.
For intermediate values of $(\beta, p)$, instead, the dynamics is mixed: initially dominated by simple contagion but following complex contagion at large times. Finally, for large $(\beta, p)$ parameter values the spreading is dominated by the simple contagion mechanism. We proposed a criterion to determine the phases of the system, confirmed by analytical expressions for one of the boundaries and by numerical methods for all of them.

Future work should be dedicated to extending the analytical study to larger $z$ values, to find a critical adoption probability $p_c$ separating simple and mixed contagion in these cases.
Furthermore, we explored the effect of heterogeneity of nodes ---with respect to their activity rate--- only qualitatively, by comparing contagion curve profiles between constant and power-law activity distributions. 
Extending the analytical and numerical frameworks presented here to the case of heterogeneous networks represents a significant avenue for future research.

\emph{Acknowledgments.}
 R.P.-S. acknowledges financial support from project PID2022-137505NB-C21, funded by MICIU/AEI/10.13039/501100011033, and by “ERDF: A way of making Europe”. M.K. acknowledges funding from the National Laboratory for Health Security (RRF-2.3.1-21-2022-00006); the DATAREDUX (ANR-19-CE46-0008); the SoBigData++ H2020-871042; and the MOMA WWTF projects.

\bibliography{main}

\clearpage
\newpage

\onecolumngrid

\begin{center}
    \textbf{\Large Supplemental Material}
\end{center}

\section{Mean-field rate equation}
\label{analytical_study_si}

We consider the time evolution of the number of infected individuals $I(t)$. 
In a microscopic time step $\Delta t$, we choose a node at random (the ego node), which is susceptible with probability $(N - I)/N$. 
The ego node becomes active with probability $a$, in which case with probability $p$ it follows a simple contagion, and with probability $1-p$ complex contagion to potentially become infected. 
Therefore, within a mean-field approximation~\cite{perra2012activity}, the number of infected nodes at time $t + \Delta t$ can be written as
\begin{equation}
    I (t+\Delta t) = I(t) + \frac{N - I(t)}{N} a \left[ p \ps  + (1-p) \pc \right],
    \label{eq:full_eq_discrete}
\end{equation}
where $\ps$ and $\pc$ are the probabilities that the focal node becomes infected by the simple and complex processes, respectively. 
For the simple contagion process, the ego node is connected to $m$ other nodes, each one infected with probability $\rho$. 
Each infected neighbor transmits the infection to the ego node with probability $\beta$. Thus, the probability that any one of the infected neighbors infects the ego is
\begin{equation}
    \ps = 1-(1-\rho\beta)^m.
    \label{eq:simple_contagion_prob}
\end{equation}
On the other hand, for complex contagion to take place, we need the ego to be neighbor of at least $\mphi$ infected nodes, an event that happens with probability
\begin{equation}
    \pc = \sum_{n=\mphi}^{m} \binom{m}{n}
   \rho^n (1 - \rho)^{m-n} \equiv \mathrm{I}_\rho(z, m+1-z),
   \label{equation_cp}
\end{equation}
where $\mathrm{I}_x(a,b)$ is the regularized incomplete beta function~\cite{abramovitz}.

Assuming that the time interval $\Delta t = 1/N$, in such a way that a whole update of the network corresponds to one Monte Carlo time step, we can take the thermodynamic limit $N\to\infty$ in Eq.~\eqref{eq:full_eq_discrete} to write the differential rate equation
\begin{equation}
    \frac{d \rho}{d t} = a (1 - \rho) \left[ p \times \ps + (1-p) \times \pc\right].
    \label{general_equation_on_rho}
\end{equation}

In the following, we analyze the beginning of the propagation process, in the limit $t \to 0$, $\rho \ll 1$, considering different cases.

\section{Pure complex contagion}
\label{seq:complex_contagion}

The case when all nodes follow the complex contagion mechanism corresponds to $p=0$.
In this scenario, using Eqs~\eqref{general_equation_on_rho} and \eqref{equation_cp}, we obtain
\begin{equation}
    \frac{d \rho}{d t} = a (1 - \rho)  I_\rho(z, m+1-z).
    \label{only_cp}
\end{equation}
Since we are interested in the behavior for small $t$ and $\rho$, we can use the power expansion of the regularized incomplete beta function for integer $z$~\cite{abramovitz}
\begin{equation}
    I_\rho(z, m+1-z) \sim \rho^z
    \binom{m}{z} , \quad \rho \to 0.
    \label{eq:binom_eq}
\end{equation}
Thus, keeping only the leading terms, Eq.~\eqref{only_cp} can be written as
\begin{equation}
    \frac{d \rho}{d t} =  C \rho^{\mphi}
\end{equation}
where the constant $C$ is equal to
\begin{equation}
    C = a \binom{m}{z}
\end{equation}

The solution of this equation, in terms of the initial density of infected seeds $\rho_0$, is
\begin{equation}
    \rho^{\mphi-1}(t) = \frac{1}{\rho_0^{1 - \mphi} - (\mphi -1) C t}.
\end{equation}
This solution shows a linear decreasing behavior in time of the function $1/\rho^{z-1}$, with a divergence at a time
$t = \rho_0 ^{1-\mphi} / [  C (\mphi - 1)]$. The time of the divergence is smaller when $\rho_0$ is higher, as initially there are more infected nodes, and when $\mphi$ is smaller, as the condition to be infected is easily reached.

The fact that the density of infected nodes exhibits a singularity at the small time description can be interpreted as a signature of the complex contagion mechanism, which usually proceeds in cascades in which a large fraction of nodes become infected in a very short period of time~\cite{watts2002simple}. This behavior is in contrast with simple contagion, which at short times is defined by a monotonous exponential increase.

\section{Mixed simple and complex contagion}
\label{Mixed_simple_and_complex_contagion}

We consider now the case of mixed simple and complex contagions. To simplify our calculations, we focus here on the case of $\mphi = 2$. 
We thus have, for the simple contagion infection probability,
\begin{equation}
    \ps = 1 - (1-\rho \beta)^m \simeq m \beta \rho - \rho^2 m (m-1) \frac{\beta^2}{2},
\label{proba_si}
\end{equation}
where we have kept only the lower order terms in $\rho$. For the complex contagion probability, from Eq.~\eqref{eq:binom_eq},
\begin{equation}
    P_2(\rho)  \simeq \rho^2 \binom{m}{2} = 
        \rho^2 \frac{m(m-1)}{2},
    \label{proba_cp}
\end{equation}
where again we have kept only the leading terms in $\rho$. Inserting Eqs.~\eqref{proba_si} and~\eqref{proba_cp} in Eq.~\eqref{general_equation_on_rho}, we obtain
\begin{equation}
    \frac{d \rho}{dt} = a p m \beta (\rho + B \rho^2)
    \label{rho_simpler}
\end{equation}
up to order $\rho^2$, and where we have defined the constant
\begin{equation}
    B = \frac{(1 - p) (m - 1)}{2 p \beta} - \frac{(m - 1) \beta}{2} - 1.
\end{equation}
The solution of Eq.~\eqref{rho_simpler} in terms of the initial density of infected nodes is
\begin{equation}
    \rho(t) = \frac{1}{A e^{-t a p m \beta} - B},
    \label{eq:solution_full_mixed}
\end{equation}
where we have defined the constant
\begin{equation}
    A = B + \frac{1}{\rho_0}.
\end{equation}

If $B$ is negative (and $\rho_0$ sufficiently small in such a way that $A > 0$), then $\rho(t)$ grows at short times, until it saturates to the value $1/|B|$.
If $B$ is positive, on the other hand, $\rho(t)$ diverges at the time
\begin{equation}
    t_\mathrm{casc} = \frac{1}{a p m \beta} \ln\left(\frac{A}{B} \right) ,
\end{equation}
which serves as a proxy to indicate the time of the cascade, namely the time of a sudden increase of the density of infected nodes. 

These different behaviors, depending on the sign of $B$ can be understood as follows. For $B<0$, the second term on the r.h.s of Eq.~\eqref{rho_simpler} is negative, in agreement with the second order expansion of the rate equation for pure simple contagion (see Eq.\eqref{proba_si}). Otherwise, if $B>0$, the  second term on the r.h.s  of Eq.~\eqref{rho_simpler} is positive, as corresponding to the pure complex contagion rule (see Eq.~\eqref{proba_cp}). We can interpret this as the value $B=0$, corresponding to the probability
\begin{equation}
    p_c(\beta) = \frac{m-1}{2 \beta + (1+\beta^2)(m-1)},
\end{equation}
as the boundary separating a dominating simple contagion (for $p > p_c$) from a dominating complex contagion (for $p < p_c$). Within the complex contagion dominated phase, the divergence time $t_\mathrm{casc}$ is a proxy of the time at which complex contagion takes over from the simple contagion prevalent at very short times.

The divergence time $t_\mathrm{casc}$ diminishes as $\rho_0$ increases, since the increase of the proportion of infected nodes occurs earlier when there is initially more infected nodes. The parameter $\beta$ also makes the critical time decreasing when it is increasing: as there is a higher probability to be infected for the simple contagion, the number of infected nodes is greater, thus the increase of $\rho$ is earlier.

In the complex contagion dominated region with $B > 0$, another way to estimate when this dynamics takes over simple contagion emerges from the analysis of Eq.~\eqref{rho_simpler}. In this equation, we have a linear term, arising from simple contagion, and a quadratic term, with components from the complex and simple contagions. Assuming that complex contagion dominates over simple contagion when the second order term becomes larger than the linear one, we can define the threshold density of infected nodes 
\begin{equation}
    \rho_\mathrm{eq} = \frac{1}{B},
    \label{rho_rho_square}
\end{equation}
such that at the time $t_\mathrm{eq}$, corresponding to $\rho(t_\mathrm{eq}) = \rho_\mathrm{eq}$, the second order term overcomes the first order one, and we expect complex contagion to prevail. From the solution Eq.~\eqref{eq:solution_full_mixed}, we can estimate, within small time approximation,
\begin{equation}
    t_\mathrm{eq} = \frac{1}{a p m \beta} \ln\left(\frac{A}{2 B} \right) 
\end{equation}
such that $t_\mathrm{eq} < t_\mathrm{casc}$. In the case that $\rho_0 > \rho_\mathrm{eq}$, the quadratic term containing elements from the complex and simple contagions dominates from the initial instant of the dynamics. This scenario corresponds to $t_\mathrm{eq} = 0$ and marks the region below which the complex contagion is fully predominant from the early stage of the dynamics. 

To sum up the behavior of $z=2$, for $p > p_c(\beta)$, contagion is dominated by the simple mechanism. For $p<p_c(\beta)$, contagion is dominated, at large times by complex the complex mechanism. At short times, however, simple contagion is prevalent, since it is a larger average transmission. At the time $t_\mathrm{eq}$, complex contagion takes over from the initially predominant simple mechanism.

\section{General case $z>2$}
\label{general_case_large_z}

We can also calculate $\rho_{eq}$ when $z > 2$.
To do so, we first aim to calculate the expression of $P(n\geq \mphi)$ and $\Delta_{m, \beta}(\rho)$ in the general case in which $\mphi$ can take any value. We first prove that $P(n\geq \mphi)$ is a polynomial for which the $\mphi^{th}$ order Taylor expansion has a lower degree of $\mphi$, meaning that $P(n\geq \mphi)$ is governed by the term $\rho^{\mphi}$. Indeed, the Taylor expansion of the term $(1 - \rho)^{m-n}$ in Eq.~\eqref{equation_cp} gives the following.
\begin{equation}
    P(n\geq \mphi) (\rho) = \sum_{n=\mphi}^{m} {m \choose n} \rho^n \left[ 1 + \sum_{i=1}^{\mphi} \frac{(-1)^i \rho^i}{i!} (m-n) \dots (m-n-i+1) \right] 
    \label{e1}
\end{equation}

In order to prove that the lowest degree of $P(n\geq \mphi)$ is $\mphi$, we consider the term $\rho^{\alpha}$, with $\alpha<\mphi$ and demonstrate that its coefficient $C_\alpha$ is null. From Equation \ref{e1}, if $\alpha$ is equal to 0, the coefficient is 1 - ${m \choose 0} \rho^0$, which is null, otherwise the coefficient $C_{\alpha}$ is the following.

\begin{equation}
    C_{\alpha} = - {m \choose \alpha} - \sum_{n=0}^{\alpha-1} {m \choose n} \frac{(-1)^{\alpha-n}}{(\alpha - n)!} (m - n) \dots (m - \alpha + 1)
    \label{e2}
\end{equation}

By reorganising the terms, Eq.~\eqref{e2} is then equal to the following.

\begin{equation}
    C_{\alpha} = - {m \choose \alpha} \left[ -1 - \sum_{n=0}^{\alpha-1} {\alpha \choose n} (-1)^{\alpha - n} \right]
    \label{e3}
\end{equation}

The right term of Eq.~\eqref{e3} can be expressed as $ -1 - \left[ \sum_{n=0}^{\alpha} {\alpha \choose n} (-1)^{\alpha - n} 1^{n} - 1 \right] $. Using the Newton's binomial, we show that that term is null, thus $C^{\alpha}$ is also null, which demonstrate that the $\mphi^{th}$ order Taylor expansion of the polynomial $P(n\geq \mphi)$ has a lower degree of $\mphi$. The coefficient of the term in $\mphi$ is then the following:

\begin{equation}
    C_{\mphi} =  - \sum_{n=0}^{\mphi-1} {m \choose n} \frac{(-1)^{\mphi-n}}{(\mphi - n)!} (m - n) \dots (m - \mphi + 1)
    \label{e2}
\end{equation}

By reorganising the term, we prove that $C_{\mphi} = {m \choose \mphi}$, and then $P(n\geq \mphi) = {m \choose \mphi} \rho^{\mphi}$. \\

The general term of $\Delta_{m, \beta}(\rho)$ is given by its Taylor expansion:

\begin{equation}
    \Delta_{m, \beta}(\rho) = - \sum_{i=1}^{\mphi} {m \choose i} (-1)^i \beta^i \rho^i 
    \label{e2}
\end{equation}

We then use those expressions in Eq.~\eqref{general_equation_on_rho}. In the right part of the equation, the term in $\rho^{\mphi}$ is $a {m \choose \mphi} [p (-\beta)^{\mphi - 1}[\beta + \frac{\mphi}{m - \mphi + 1}] + 1 - p] \rho^{\mphi} $ and the term in $\rho$ is $- a p m \beta \rho$. Those two terms are equivalent when:

\begin{equation}
\rho_{eq} \approx \sqrt[\mphi - 1]{\frac{m p \beta}{{m \choose \mphi} [p (-\beta)^{\mphi - 1}[\beta + \frac{\mphi}{m - \mphi + 1}] + 1 - p]}}
\label{rho_equ_general_case}
\end{equation}

\section{Comparison of different activity distributions}

\begin{figure*}[htbp]
    \centering
    \includegraphics[width=0.9\columnwidth]{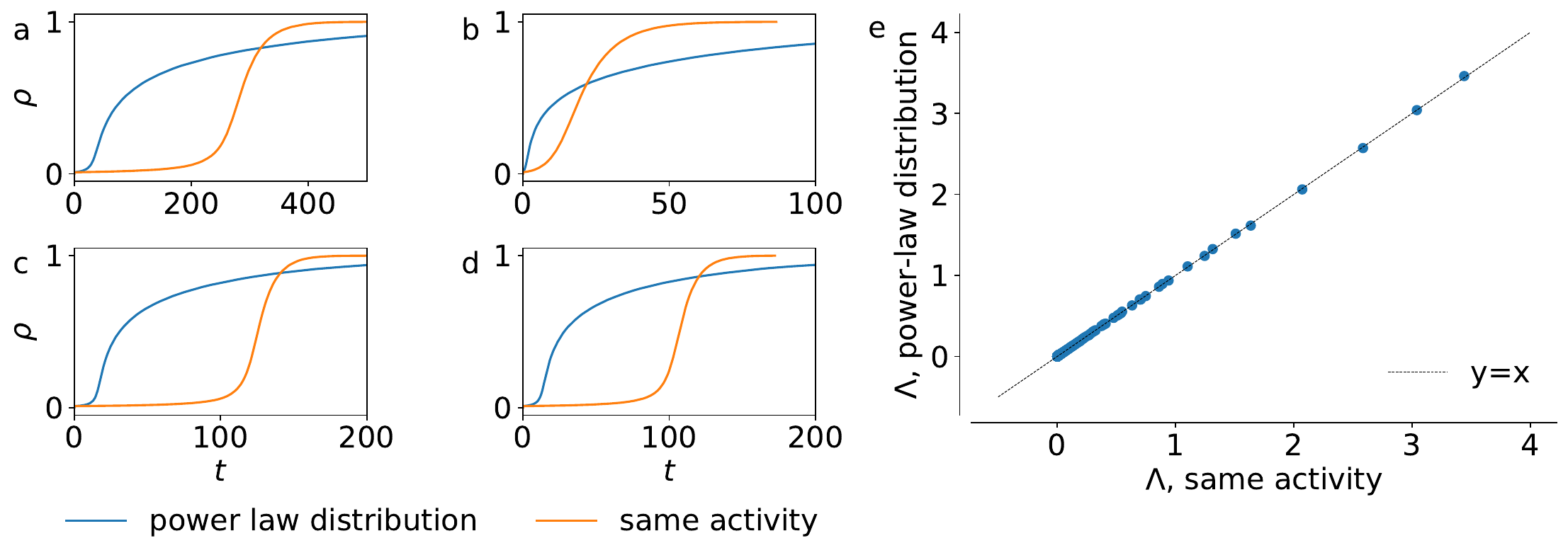}
       \caption{Comparison of the contagion curves (panels a to d) and the final ratio $\Lambda$ (panel e) when every node has the same activity and when the activities are sampled from a power law. Both distributions have the same average, i.e., 0.1. The exponent of the power law distribution is 1.14. The simulations on the left part are parameterised with $(\beta, p)$: a - $(0.05, 0.99)$, b - $(0.99, 0.99)$, c - $(0.05, 0.05)$, and d - $(0.99, 0.05)$. The propagation is considerably slower when the nodes have the same activity, but reaches faster the final state. On panel e, every scatter point stands for a couple of parameter $(\beta, p)$. The final ratio $\Lambda$ is identical for the two activity distributions.
    }
    \label{fig_comparison_activities}
\end{figure*}

We aim to compare the results of two activity distributions: a dirac distribution (every node has the same activity, used in the main manuscript) and a power law distribution. 
Both distributions have the same average of 0.1, thus the exponent of the power law is 1.14. 
We observe that the simulations made which the dirac distribution are at first slower than the ones made with a power law distribution, but reaches faster the final state (Figs.~\ref{fig_comparison_activities}, panels a to d).
Indeed, as some nodes have a high activity with the power-law distribution, they are infected first and accelerate the beginning of the process. 
However, as the majority nodes have a low activity, the process is slow to reach the contagion of all the nodes. 
Despite of the difference of celerity, the outcome of the two distributions are identical: the final ratio between the number of nodes infected by the simple and the complex contagions is the same (Fig.~\ref{fig_comparison_activities}e).

\section{Simulations on the extended parameter-space $(\beta, p)$}
\label{extended_parameter_space}

\begin{figure*}[htbp]
    \centering
    \includegraphics[width=0.95\textwidth]{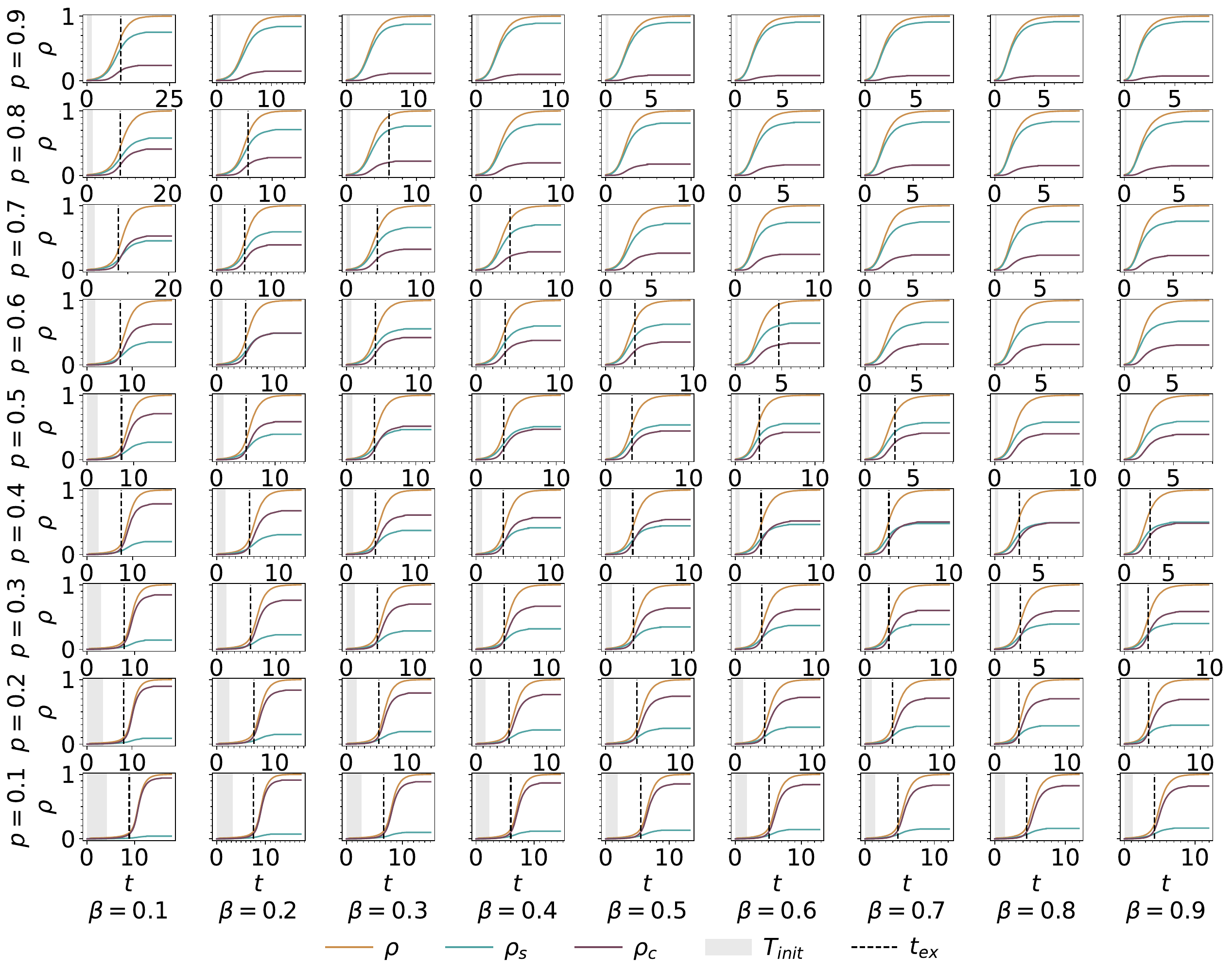}
    \caption{Fraction of infected nodes, $\rho(t)$, as the function of time for simulated spreading scenarios (yellow line), together with the proportion of nodes infected by the simple, $\rho_s$, (blue curve) and the complex, $\rho_c$, (purple curve) contagions for $z=2$. Panels show simulation results averaged over $100$ realisations, for different values of $\beta$ (x-axis) and $p$ (y-axis).
    }
    \label{fig_rho_s_rho_c_general}
\end{figure*}

The contagion curves and the expected times of increase $t_{casc}$, displayed on Fig.~2 in the main text for four different parametrisations $(\beta, p)$, are shown for a larger range of values on Fig.~\ref{fig_rho_s_rho_c_general}. We can observe that large values of $p$ and small values of $\beta$ lead to a slow dynamics, as the model attempts but fails to infect nodes through the simple contagion. In contrast, the fastest contagion processes are for high values of both $p$ and $\beta$ as the nodes in this setting are mainly infected successfully by the simple contagion.

On the same figures, we also explore the proportion of nodes infected by each process through time, namely $\rho_s$ and $\rho_c$. 
Complex contagion dominates the propagation for small values $p$, as nodes are more likely infected by the complex contagion in this setting 
Also, in line with previous results, the simple contagion governs the spreading when $p$ is high.
The influence of $\beta$ is minor but observable, leading to higher $\rho_s$ when $\beta$ increases. In particular, in the simulations for $p=0.5$, $p=0.6$ and $p=0.7$, raising $\beta$ changes the contagion process dominating the simulation, from the complex to the simple contagions.

\begin{figure*}[tbp]
    \centering
    \includegraphics[width=0.95\textwidth]{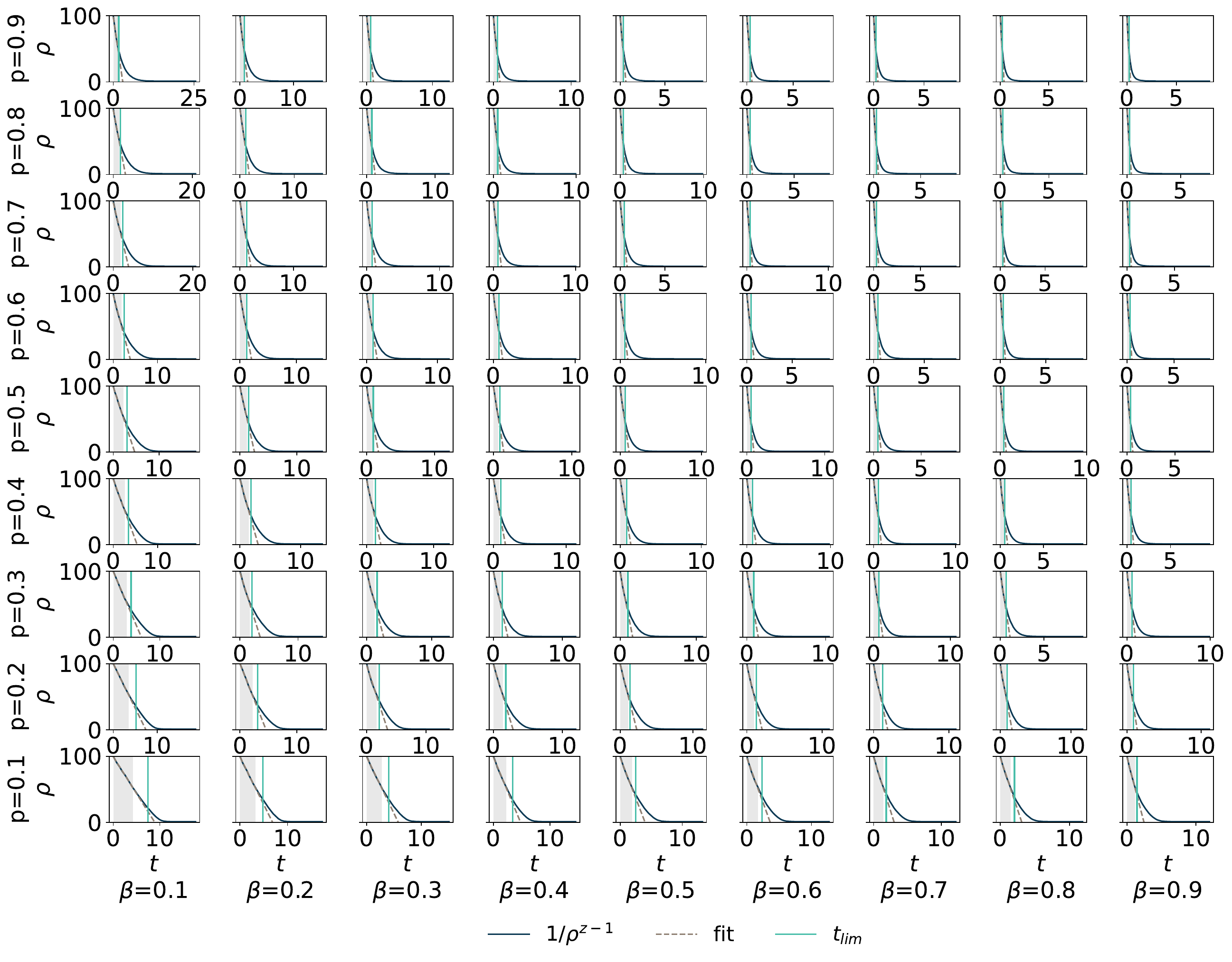}
    \caption{Inverse of the proportion of infected neighbors to the power $\mphi$-1 (dark blue line), fitted with a linear function on the first part of the propagation (black dotted line). The x-axis stands for different values of $\beta$, while the y-axis represents the values of $p$.The time $t_{lim}$ when the difference between $1/\rho^{\mphi}$ and its fit is higher than $\epsilon>5$ is indicated with a light blue line.}
    \label{fig_linear_general}
\end{figure*}

Fig.~\ref{fig_linear_general} exemplifies the second method to evaluate the dominance of the simple or complex contagions at the beginning of the process with an extended parameter-space compared to Fig.~3 in the main text. The curve $\/\rho^{\mphi - 1}$ is linear decreasing if the complex contagion dominates, which is the case for low values of $\beta$ and $p$. We measure the linearity of the early times of the curves by fitting the curves with a line and showing the time when the simulations and the fits have a difference higher than $\epsilon$. As expected, the linear trend is not present for high values of $\beta$ and $p$, for which the simple contagion governs.

\clearpage

\section{Supplemental Figures}

\end{document}